\documentclass{icrc2009}

\usepackage{graphicx}   
\usepackage[caption=false]{caption}    
\usepackage[font=footnotesize]{subfig} 
\usepackage{fixltx2e}
\usepackage{url}

\newcommand{\shorttitle}[1]%
{\markboth{Proceedings of the 31\MakeLowercase{$^{st}$} ICRC, {\L}\'{o}d\'{z} 2009}{#1} }

\newcommand{\pubjournal}[6] {#1, #2 {\bf #3}, #4 (#5).}

\begin{document}
\title{The GALPROP Cosmic-Ray Propagation Code  }

\author{\IEEEauthorblockN{A. W. Strong \IEEEauthorrefmark{1}, 
			  I. V. Moskalenko\IEEEauthorrefmark{2},
                          T. A. Porter\IEEEauthorrefmark{3},
                          G. J\'ohannesson\IEEEauthorrefmark{2},
                          E. Orlando\IEEEauthorrefmark{1},
                          S. W.  Digel\IEEEauthorrefmark{2} 
   }
                            \\
\IEEEauthorblockA{\IEEEauthorrefmark{1}Max-Planck-Institut f\"ur extraterrestrische Physik, Garching, Germany}
\IEEEauthorblockA{\IEEEauthorrefmark{2}Hansen Experimental Physics Laboratory and Kavli Institute for Particle  Astrophysics and Cosmology,\\ Stanford University, Stanford, CA 94304, USA}
\IEEEauthorblockA{\IEEEauthorrefmark{3}Santa Cruz Institute for Particle Physics, University of California, 1156 High Street, Santa Cruz, CA 95064, U.S.A.}
}

\shorttitle{Strong et al. GALPROP code}

\maketitle

\begin{abstract}
The Galactic cosmic-ray propagation code GALPROP is designed to make
predictions of many kinds of data self-consistently, including direct
cosmic-ray measurements, gamma rays and synchrotron radiation. In the
decade since its conception it has undergone considerable development,
which is continuing. A public version was made available a few years ago,
supported by a website. We describe the new
features of the current version which will become part of the next public
release. Plans for future developments are also mentioned.
\end{abstract}


\section{Introduction}
The GALPROP code \cite{SM98}
has now 15 years of development behind it. Originally written in fortran90, a completely new version in C++ was produced in 2001.
GALPROP is designed to make predictions of many kinds of observations: cosmic-ray (CR) direct measurements including primary and secondary nuclei, electrons and positrons, gamma rays, synchrotron radiation, etc. 
The important concept is that these various 
data are all related to the same astrophysical components of the Galaxy and hence have to be modelled self-consistently. 
GALPROP is foremost a CR  propagation code, solving the relevant equations numerically in 2D or 3D. 
The goal is for GALPROP-based  models to be as realistic as possible and to make use of  available astronomical information, with a minimum of  simplifying restrictions.
A complete description of the rationale and  motivation  is given in the review \cite{AnnRev57}.
More details are in \cite{SMR00,Moskalenko2003,SMR04,SMRRD04}.

The code solves the transport equations numerically on a 2D or 3D grid, accounting for nuclear spallation processes and energy losses.  
The solution is time-dependent; normally it is continued until a steady-state is reached, but the explicit solution in time is also useful for studying the effect of discrete stochastic sources.
After the CR propagation has been computed for all species including secondaries, the CR spectrum at each point in the Galaxy
is used to compute gamma rays using gas surveys and a model of the interstellar radiation field, ensuring a consistent approach.
Synchrotron emission is also computed using the electron and positron spectra and a 3D model of the Galactic magnetic field.
The main parameters for a given GALPROP model are the CR primary injection spectra,
the  spatial distribution of CR sources, the size of the propagation region, the spatial and momentum diffusion coefficients and their dependencies on particle rigidity. 
The interstellar gas distribution is based on observations of the neutral atomic and molecular gas, ionized gas,  while the interstellar radiation field (ISRF) is
based on a new calculation. 
The Galactic magnetic field model can be chosen from among  various models taken from the current literature, suitably parameterized to allow fitting to the observations.

The GALPROP code
computes a complete network of primary, secondary and tertiary CR
production as described in \cite{SM98,SMR04,Ptuskin06}, starting from
input source abundances.
The code includes cross-section measurements and energy-dependent 
fitting functions \cite{SM01,Moskalenko2003}.
The nuclear reaction network is built using the
Nuclear Data Sheets. The isotopic cross-section database
is built using the extensive T16 Los Alamos compilation
of the cross-sections \cite{Mashnik1998} and modern
nuclear codes CEM2k and LAQGSM \cite{Mashnik2004}. 
The most important isotopic production cross-sections
(2H, 3H, 3He, Li, Be, B, Al, Cl, Sc, Ti, V, and Mn)
are calculated using our fits to major production channels
\cite{Moskalenko2001,Moskalenko2003}. 
Other cross-sections are calculated using
phenomenological approximations by Webber et al. \cite{Webber1990}
and/or Silberberg and Tsao \cite{ST1998}  renormalized to
the data where they exist.

The propagation equation is solved numerically starting
at the heaviest nucleus (e.g.  $^{64}$Ni), computing all the resulting
secondary source functions,  then proceeding to the
nuclear species with $A-1$. The procedure is repeated down to
$A=1$. To account for some special $\beta^-$-decay cases (e.g.,
$^{10}$Be$\to^{10}$B) the whole loop is repeated twice. 
While for many purposes  the
2D cylindrically symmetrical option is sufficient, the 3D case is more realistic and will be the main option used in future.
The full time-dependent 3D solution is required for studying the effect of stochastic (in space and time) sources,
as discussed  in \cite{Strong2001}.

\section{Management and community use}
GALPROP is a public code but is
in continuous development by a small team.
It is developed and maintained using a variety of open source tools.
Revision control uses subversion\footnote{http://subversion.tigris.org}.
Configuration and installation management is via the gnu
autotools\footnote{http://www.gnu.org}. The code has been successfully
installed under a variety of unix-like systems, including various Linux
distributions and versions of the Mac OSX system, using different versions
of the gcc  and intel compilers.

A website\footnote{http://galprop.stanford.edu} with a forum for discussion of all related aspects is maintained, with currently about 90 registered users.

A considerable number of papers in the literature each year make use of GALPROP.
A principal current application of GALPROP is as the reference model for the Galactic diffuse gamma-ray emission and CR electrons for the LAT instrument on NASA's Fermi Gamma-Ray Observatory \cite{Porter2009,Abdo2009,Strong2009}.
Other mission-related applications are CR for PAMELA, gamma rays for INTEGRAL,  and synchrotron predictions for Planck.
A GALPROP-based synchrotron study can be found in \cite{Orlando2009}.

\section{New features}
Features which have been added since the last release and which will appear in the next release include:

\noindent$\bullet$  A new calculation of the Galactic ISRF  using the {\em FRaNKIE\footnote{Fast Radiation transport Numerical Kode for Interstellar Emission}} code\cite{Porter2008}.

\noindent $\bullet$ HEALPix\footnote{http://healpix.jpl.nasa.gov} output of gamma-ray and synchrotron skymaps. HEALPix \cite{Gorski2005} with its uniform sky coverage, equal-area pixels and powerful functions (convolution, harmonic analysis) is a standard for radio-astronomy applications and WMAP, Planck etc. Therefore it is an advantage to have this format as well as the more conventional (l,b) grid.

\noindent$\bullet$  MapCube output for compatibility with Fermi-LAT Science Tools software\footnote{http://fermi.gsfc.nasa.gov/ssc/data/analysis}.

\noindent$\bullet$  Gamma-ray skymaps output in Galactocentric rings to facilitate spatial analysis.

\noindent$\bullet$ Implementation of alternative treatments of pion production.

\noindent$\bullet$  3D modelling of the Galactic magnetic field, regular and random, with a range of models from the literature, extensible to any new model as required.

 \noindent$\bullet$ 
Calculations of  synchrotron skymaps on a frequency grid, using both regular and random magnetic fields, and the CR electron and positron spectra as a function of position (no approximating power-law assumption).

\noindent $\bullet$ Parallel support with openmp

\noindent$\bullet$  Memory usage optimization.

\noindent$\bullet$  New HI and CO survey data, with more precise assignment to Galactocentric rings.

\newpage 

\noindent $\bullet$ More accurate line-of-sight integration for computing gamma-ray skymaps.

\noindent$\bullet$  Anisotropic inverse Compton scattering calculations efficiently implemented

\section{New release}
The last public release was in 2004, and now  a new release is planned to coincide with this ICRC.
It will be made available via the GALPROP website cited above.


\section{Future developments}
While GALPROP represents the current state-of-the-art in its field, it has definite limitations, partly historical.
For this reason we plan a completely new development.
Features will include an adaptive grid.
In addition we will adopt a more modern code structure taking full advantage of C++ and a modular design.
This will mean that GALPROP functions will be usable by other codes, and new GALPROP-like codes can be built using the libraries provided at the users' wishes.
Suggestions for this development from the CR community via the GALPROP forum are invited.

\section{Acknowledgments}

I. V. Moskalenko and T. A. Porter acknowledge support from NASA
grant NNX09AC15G.

\end{document}